\documentclass[journal,twoside,web]{ieeecolor}
\usepackage{generic}
\usepackage{amsfonts}
\usepackage{graphicx}
\usepackage{textcomp}
\usepackage{amsmath}
\usepackage{cite}
\usepackage{amssymb}
\usepackage{bm}
\usepackage{epstopdf}
\usepackage{caption}
\usepackage{newunicodechar}
\newunicodechar{́}{\'} 

\usepackage{multirow}
\usepackage{array}
\usepackage{booktabs}
\usepackage{threeparttable}
\usepackage{subfigure}
\usepackage{float}
\usepackage{diagbox}
\usepackage{makecell}
\usepackage{algorithm}
\usepackage{algorithmic}
\usepackage{amssymb}
\usepackage{color}
\usepackage{setspace}

\usepackage[dvipsnames]{xcolor}
\usepackage{pdfpages}

\usepackage{tikz}
\usepackage{hyperref}
\usepackage{cleveref}
\hypersetup{hidelinks}

\def\BibTeX{{\rm B\kern-.05em{\sc i\kern-.025em b}\kern-.08em T\kern-.1667em\lower.7ex\hbox{E}\kern-.125emX}}

\begin{document}
	
	\title{\huge A Sampling Model for Grid Material Inspection Based on Analytic Hierarchy Process with Absolute Measurement}
	\author{\IEEEmembership{Jing Xu and Yongbo Zhang}
	\thanks{All authors are independent researchers. \itshape  \upshape}}
	\maketitle

	\begin{abstract}
		The power grid is a fundamental industry that is vital to the national economy and people's livelihoods. Ensuring the safety and reliability of the power grid is an everlasting concern for power grid enterprises, and it requires high-quality power grid materials to guarantee secure and reliable power supply. As the demand for reliable and safe grid materials increases, effective inspection methods are crucial for ensuring the quality and compliance of these materials. This paper proposes a sampling model for inspecting grid materials based on improved Analytic Hierarchy Process (AHP). The model extends the traditional AHP framework and incorporates improvements to address specific requirements for grid material inspection. The objective of this model is to enhance the accuracy and effectiveness of the detection process while considering the complex decision-making involved in material selection. The method selects characteristic variables of specific material equipment and calculates the weights of these variables based on AHP. By utilizing historical sampling data from the Enterprise Control Platform (ECP), the levels of inspection results are determined, and weighted comprehensive performance scores for each piece of material equipment are calculated. Subsequently, the equipment is sampled and tested according to the desired sampling scale, with sorting based on the ascending order of comprehensive performance values. To demonstrate the effectiveness of the proposed model, a case study involving grid material inspection is presented. The results indicate that compared to random sampling methods, the proposed model greatly improves the accuracy and efficiency of grid material testing.
	\end{abstract}
	
	\begin{IEEEkeywords}
		Sampling model, Material inspection, Analytic Hierarchy Process, Quality control, Decision-making
	\end{IEEEkeywords}

	\section{Introduction}
	The power grid plays a critical role in modern society as it serves as the backbone of electrical infrastructure, facilitating the reliable and efficient transmission and distribution of electricity\cite{li2017research, weaver2016cyber}. The power grid is a vital system that enables the functioning of various sectors, including residential, commercial, industrial, and governmental operations. Consequently, power grid security is an eternal concern for electric grid enterprises. To ensure a secure and reliable supply of electricity, high-quality grid materials are indispensable.
	
	The equipment and materials within the power grid function as conduits for transmitting electrical energy. Strengthening the quality supervision system for grid materials is a foundational task to ensure the high quality of these materials. The quality testing of power equipment plays a pivotal role in power material management, grid security, and operational maintenance. Power grid companies annually procure a substantial quantity of distribution network materials. Consequently, conducting pre-installation sampling inspections on these materials is a crucial step in ensuring the quality control of grid materials and maintaining the safe and stable operation of the distribution network.

    At present, there is a significant annual procurement of materials for distribution network construction. The increasing technological
complexity and sophistication of some equipment lead to rising inspection costs and longer inspection times. To ensure the quality control of purchased distribution network materials, power grid companies commonly employ sampling inspections. The current sampling standard for distribution network material inspection in the power grid relies on a percentage-based random sampling approach. Irrespective of batch size, the same acceptance number is specified, and samples are randomly selected in the same proportion from the product batch. While this method boasts a simple and easy-to-implement process, it falls short in guaranteeing the accuracy of inspection results when dealing with small sample sizes. Increasing the sampling ratio would escalate the cost and workload of inspections. In the procurement of distribution network materials, many critical materials have relatively small batch sizes, posing a typical small-sample sampling problem. Therefore, in small-sample sampling inspections, it is imperative to adhere to the requirements outlined in power grid quality supervision documents while ensuring scientific rigor and cost-effectiveness, presenting an urgent issue that needs to be addressed.
	
	The AHP\cite{saaty1988analytic} is a decision-making method that involves decomposing complex decision elements, such as objectives, criteria, and alternatives, into hierarchical levels for qualitative and quantitative analysis. This method originated for the U.S. Department of Defense on the topic of `electric power allocation based on the contribution of various industrial sectors to national welfare. AHP is favored for its capacity to handle limited sample sizes, maintains a consistent approach, provides simplicity, and offers user-friendly software options as key justifications for its utilization\cite{darko2019review}.

	The existing sampling standard for equipment detection is percentage random sampling. Although the process of this method is simple and easy to implement, it cannot guarantee the accuracy of small sample sampling, nor can it control the cost of large sample sampling. In order to screen out more representative equipment in small sample detection, this paper proposes a sampling model for grid material inspection based on improved AHP. This method selects characteristic variables of specific material equipment, calculates the weights of these variables based on AHP, determines inspection result levels based on historical sampling data from the Enterprise Control Platform (ECP), and finally calculates the weighted comprehensive performance scores of each piece of material equipment in the same batch. The equipment is then sampled for testing in ascending order of the comprehensive performance values, according to the required sampling size. By scientifically and accurately assessing the quality of the tested products and selecting materials with poorer comprehensive performance for inspection, the material testing sampling model can balance the scientific rigor and inspection cost, improve quality and efficiency, and effectively fulfill the quality supervision role of sampling inspections. It plays a crucial role in quality control before equipment installation. The main contributions of this paper are summarized as follows.
    \begin{itemize}
\item Our method makes full use of historical sampling data and historical testing results, and uses an analytic hierarchy process to build a sampling model for material testing.
\item Our method can improve the accuracy of small sample sampling, thus saving the cost and workload of sampling inspection.
\item We designed the quality scoring system of power grid materials and equipment, and realized the quality scoring of the equipment sent for inspection.
\end{itemize}

	\section{Related Work}
	
	Saaty \textit{et al.}\cite{saaty1986absolute} explored two types of measurement in the Analytic Hierarchy Process (AHP): absolute measurement and relative measurement. Absolute measurement involves comparing elements or alternatives against a standard, particularly at the lower levels of the hierarchy. This process ensures the preservation of the absolute rank of alternatives, regardless of the number introduced. De Felice \textit{et al.}\cite{de2013absolute} applied the AHP model to identify a quality model for assessing racecourses based on criteria such as quality organization of racing, infrastructure and equipment, attractiveness, and management skill. The study focuses on incorporating intangible criteria into the decision-making process and emphasizes the measurement of these criteria. Salomon\cite{salomon2016absolute} explored the reasons why absolute measurement and ideal synthesis are rarely used in the Analytic Hierarchy Process (AHP). The study pointed out the deficiency of the AHP method in this respect and put forward the possibility and potential of absolute measurement and the ideal synthesis of the AHP method.
	
	A limited number of researches on sampling inspections have been recently proposed, specifically for grid materials. Most of them focus on the application in the area of food and agricultural products. Wu \textit{et al.}\cite{WU2014239} developed a computer-aided sampling scheme for quality control in the food industry, which includes distribution identification, measurement error adjustment, and sampling inspection plan design. Sommer \textit{et al.}\cite{SOMMER201989} proposed a multistage acceptance sampling scheme for quality control of photovoltaic systems, considering error probability control, dependent sampling, and providing explicit formulas and a recursive algorithm for sampling plan calculation. Atefe \textit{et al.}\cite{BANIHASHEMI2021107155} presented a novel methodology for designing cost-effective resubmitted and multiple dependent state (MDS) sampling plans. The approach integrates the process yield index, Taguchi loss function, and the average sample size to minimize costs associated with the yield index. Julia \textit{et al.}\cite{thomas2023design} put forward the GFMDS sampling plan as a novel approach for attribute sampling, considering the ambiguity in determining the exact value of defectives. This plan incorporates fuzzy logic principles, providing a flexible and robust framework for attribute sampling in quality control.

	AHP has been gaining attention in the field of sampling methods. Traditional sampling methods typically rely on statistical theories and probability inference. However, in specific research areas or application scenarios, AHP offers a new approach for making sampling decisions and selecting samples. Do \textit{et al.}\cite{do2013calculating} employed the AHP to determine the appropriate sampling frequency for each station, as well as for the entire network, by combining weighting factors of variables and the relative weights of stations. Nikolaos \textit{et al.}\cite{apostolopoulos2016regional} investigated the relationship between regional factors and the investment attractiveness of solar energy production in Greek regions. They developed a rank order of the regions based on their investment attractiveness by using the AHP and expert sampling. Kumar \textit{et al.}\cite{kumar2017soil} utilized the AHP method to guide the selection of variables, assess their importance through pairwise comparisons, assign weights, and calculate the Soil Quality Index, providing a systematic and robust approach for soil quality assessment. Singh \textit{et al.}\cite{singh2023assessing} proposed a new water quality index method based on AHP to evaluate the inland surface water quality. The method can be applied effectively in different locations with varying parameters to evaluate surface water quality.
	
	\section{Materials and methods}
	
	\subsection{Analytic Hierarchy Process}

	The AHP approaches a complex multi-objective decision problem as a system and dissects the objectives into numerous sub-objectives or criteria. These criteria are then further subdivided into several levels of indicators, which can be either criteria or constraints. Employing qualitative indicator fuzzy quantification methods, the AHP computes hierarchical weights and overall rankings. This systematic approach provides an optimized decision-making framework for scenarios involving multiple objectives and alternatives.

	AHP can be generally divided into three levels: the project level, the criterion level, and the alternative level\cite{VAIDYA20061}. The criterion level can be further divided into multiple sub-levels, depending on the complexity of the overall objective. The main approach to constructing a corresponding AHP model is as follows: firstly, determine the overall objective problem, then refine the objective problem into different judgment criteria and different solutions.
	
	Ultimately, by considering distinct relationships, such as adjacency or hierarchical connections between criteria and solutions, these elements can be categorized into various levels: the project level, the criterion level, and the alternative level, thereby forming a hierarchical structural model. The primary aim is to decompose the ultimate problem objective into solution options at the alternative level. Within this alternative level, factors are ranked according to their relative importance or superiority concerning the project level. The establishment of an AHP model and the associated computations primarily involve the following four steps.

	\subsubsection{Establishing a hierarchical structure model}

The objective problem is systematically broken down into various objective criteria and distinct decision alternatives, progressing layer by layer through the project level, criterion level, and alternative level. The corresponding hierarchical structure model is constructed based on the hierarchical relationships among the objective criteria and decision alternatives. Within this hierarchical structure model, the project level holds the highest position, symbolizing the decision-making purpose and the intricate problem requiring resolution. Occupying the lowest tier, the alternative level encompasses all conceivable solution alternatives available to the decision-maker to achieve the objective. Positioned at the intermediate level, the criterion level includes the decision criteria, which can be further subdivided into multiple sub-levels based on the real-world context.

	\subsubsection{Constructing judgment (pairwise comparison) matrices}
	
	When constructing pairwise comparison matrices, it is necessary to consider not only the qualitative importance relationships between the constituent factors of the hierarchical structure model but also quantitatively determine the weight relationships between the factors at each level. The constructed pairwise comparison matrices possess the following property\cite{Alireza2009}.
	
	\begin{footnotesize}
		\begin{equation}
			\begin{aligned}
				a_{ij} = \frac{1}{a_{ji}}, \hspace{0.5em} a_{ji}>0,
			\end{aligned}
			\label{formula2}
		\end{equation}
	\end{footnotesize}

	\noindent where the term $a_{ij}$ represents the element at the intersection of the $i$-th row and $j$-th column in the matrix. The numerical value of $a_{ij}$ represents the relative importance assigned to the $i$-th factor compared to the $j$-th factor in the pairwise comparison according to Table \ref{Importance}.
	\vspace{0.2em}

	\begin{table}[ht]
		\renewcommand{\arraystretch}{1.3}
		\caption{The assignment method for pairwise comparison matrices in the AHP}
		\centering
		\begin{tabular}{c|c}
			\hline
			\multicolumn{1}{c|}{\multirow{1}{*}{\textbf{Relative Importance of}}} & \multicolumn{1}{c}{\multirow{2}{*}{\textbf{Quantized Value}}}\\
			\multicolumn{1}{c|}{\textbf{Factor $\bm{i}$ to Factor $\bm{j}$}}\\
			\hline
			\multirow{1}{*}{Equally Important}&1\\
			\hline
			\multirow{1}{*}{Slightly Important}&3\\
			\hline
			\multirow{1}{*}{More Important}&5\\
			\hline
			\multirow{1}{*}{Strongly Important}&7\\
			\hline
			\multirow{1}{*}{Particularly Important}&9\\
			\hline
		\end{tabular}
		\vspace{0.5em}
		
		\small{\textbf{Notes:} The median values of the two adjacent judgments can also be set as 2,4,6, and 8, respectively.}
		\label{Importance}
	\end{table}

	\subsubsection{Hierarchical single ranking and consistency test}
	
	To obtain the maximum eigenvalue $\lambda_{max}$ and its corresponding eigenvector of a pairwise comparison matrix, the eigenvector is first normalized to obtain the single-level weights for the corresponding risk elements. The maximum eigenvalue is used as a dependent variable for consistency evaluation, and the Consistency Index ($CI$)\cite{AGUARON2003137} can be calculated by
	
	\begin{footnotesize}
		\begin{equation}
			\begin{aligned}
				CI = \frac{\lambda_{max}-n}{n-1},
			\end{aligned}
		\end{equation}
	\end{footnotesize}
	
	\noindent where $n$ represents the order of the pairwise comparison matrix. When $CI$ is small, the pairwise comparison matrix meets the consistency criteria. When $CI$ is large, it indicates that the pairwise comparison matrix is inconsistent and requires appropriate adjustments to the factors. Although $CI$ can provide a measure of consistency, it cannot provide a fixed value as a basis for judgment. In order to better assess the level of consistency achieved in constructing the pairwise comparison matrix, the AHP also introduces a Random Index ($RI$)\cite{SHRESTHA2004185} as a basis for judgment.

	\begin{footnotesize}
		\begin{equation}
			\begin{aligned}
				RI = \frac{CI_1+CI_2+\cdots+CI_n}{n},
			\end{aligned}
		\end{equation}
	\end{footnotesize}
	
	\noindent where $RI$ is related to the order of the pairwise comparison matrix, which corresponds to the number of risk factors at that level. Random factors may introduce inconsistencies in the pairwise comparison matrix during the model construction process. So, it is necessary to verify the Consistency Ratio ($CR$)\cite{FRANEK2014164} even when the pairwise comparison matrix meets the consistency requirement in studies using the AHP.
	
	\begin{footnotesize}
		\begin{equation}
			\begin{aligned}
				CR = \frac{CI}{RI}.
			\end{aligned}
		\end{equation}
	\end{footnotesize}
	
	In general, if $CR$ is less than 0.1, it is considered that the pairwise comparison matrix constructed for the current level model has passed the consistency test, and the model is deemed usable. Otherwise, if $CR$ exceeds 0.1, it indicates a failure to pass the consistency test and the corresponding risk factors that did not pass the test need to be removed.

	\subsubsection{Hierarchical overall ranking and consistency test}

After deriving individual rankings for each level, it becomes imperative to establish a direct connection between the project level and the alternative level. Consequently, the overall ranking of the hierarchy must be computed. This process entails multiplying the weights of the preceding level with the weights associated with the risk factors at the alternative level. The resultant product represents the comprehensive weight, and the hierarchy is subsequently ranked based on the comprehensive weights assigned to the risk factors.

	\subsection{Analytic Hierarchy Process with absolute measurement}
	
\begin{figure}[H]
		\centering
		\includegraphics[width=0.36\textheight]{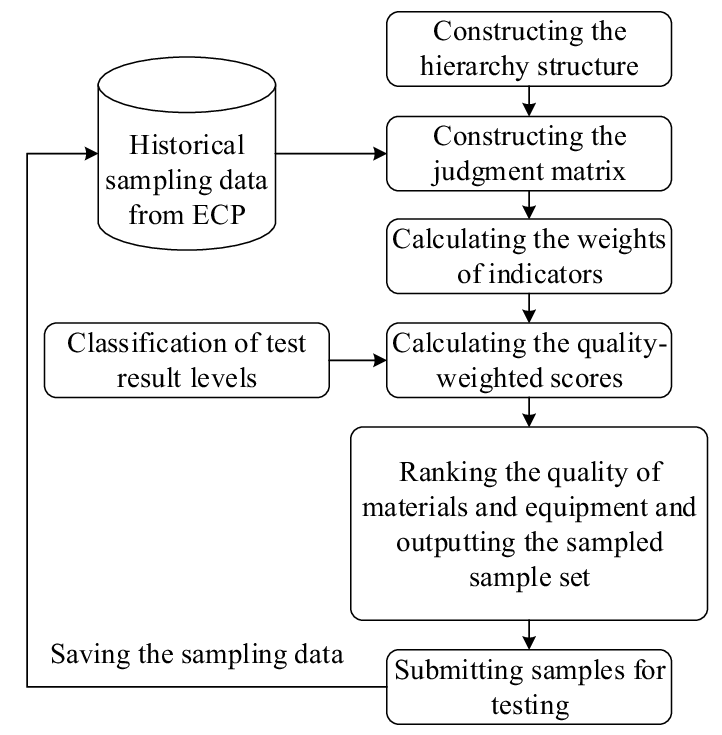}
		\caption{Algorithm flowchart of the material testing sampling model based on the AHP}
		\label{SamplingModel}
	\end{figure}

	Based on the historical sampling data of ECP and the classification of test results, a sampling model of material detection was constructed by using AHP. The model consists of six parts, and the specific implementation steps are shown in Fig. \ref{SamplingModel}.
	
	\subsubsection{Constructing the hierarchy structure}

The initial step of the AHP involves organizing the target problem into a hierarchical structure comprising the goal level, criteria level, and indicator level, creating a structured hierarchy with multiple objectives and levels. In this study, the objective is to evaluate the performance quality of materials and equipment before sampling, followed by selecting equipment with lower performance quality for testing based on the ranking. Consequently, specific statistical characteristic variables representing the performance quality level of materials and equipment need to be identified as evaluation indicators. These indicators serve as the foundation for ranking the performance quality of materials and equipment, and a hierarchical structure model is formulated utilizing these evaluation indicators.

	\subsubsection{Constructing the judgment matrix}
	The judgment matrix compares each element of the lower level with the elements of the upper level based on judgment criteria to determine the values of the matrix. It reflects the subjective understanding and evaluation of the relative importance of elements based on objective reality. The matrix elements are typically assigned values using a 1-9 scale\cite{5579387}, where the assigned values may be influenced by subjective factors to some extent. In this evaluation method, we have made improvements to the AHP. Firstly, we calculate the nonconformity rate of product testing caused by evaluation indicators using historical sampling data from ECP. Then, we use the ratio of data from two evaluation indicators as the matrix element values to objectively reflect the data value. A matrix element of 1 indicates that the two indicators are equally important, while an element greater than 1 indicates that the former is more important than the latter, and so on.
	
	\subsubsection{Calculating the weights of indicators}

	The judgment matrix is processed using the sum-product method and the square root method to obtain its maximum eigenvalue and corresponding eigenvector. The resulting eigenvector is then normalized, and the normalized vector represents the single-level weights for the corresponding risk elements. A consistency test is also conducted to ensure the rationality of the weights.

	\subsubsection{Classification of test result levels}
	The test results of the material and equipment testing items are divided into four levels: excellent, good, qualified, and basic qualified. The scores are assigned to $a$, $b$, $c$, and $d$ ($a>b>c>d$), respectively. The quality of products is distinguished according to the scores. The scores of each test item are defined as $G_i$ ($i=1,2,\cdots,n$, where $n$ is the number of test items).

	\subsubsection{Calculating the quality-weighted scores}
	According to the weight value of the material and equipment index and the corresponding test results, the quality weighted score of the sampled products is obtained through mathematical calculation. Set a conformance determination value $J$, when any test item of the product is unqualified, that is, there is $G_i=0 \hspace{0.5em} (i=1,2,\cdots,n)$, let $J$=0, otherwise $J$=1, the product mass weighted fraction $M$ is calculated by

	\begin{footnotesize}
		\begin{equation}
			\begin{aligned}
				M = J\sum_{i=1}^{n}W_i G_i.
			\end{aligned}
		\end{equation}
		\label{qualityWeight}
	\end{footnotesize}

	The improved AHP mainly calculates the ratio of each evaluation index based on objective data to construct a judgment matrix, and adopts the method of horizontal pair-to-pair comparison of the importance of evaluation indicators. The assignment results are completely dependent on objective data, which reflects the importance of detection data in product quality evaluation.

	\subsubsection{Sampling and submitting samples for testing}
	
	The weighted score of the comprehensive performance of each material and equipment in the same batch is calculated, the comprehensive performance value is sorted from small to large, and the material and equipment are extracted and sent to the sample for testing according to the sampling quantity.
	
	The sampled materials and equipment will be sent to the sample for testing, and the tested data will be stored in the ECP as the data input when constructing the judgment matrix of the same materials and equipment next time.

	\section{Results and discussion\label{experiments}}
	The selected research subject of this paper is low-voltage power cables. The analysis is based on a sample of 500 low-voltage power cable inspection reports obtained from ECP, illustrating the implementation steps of the improved AHP sampling model.

	\subsection{Test item weight assignment}
	
	When selecting test items as quality evaluation indicators, it is important to choose indicators that objectively and impartially reflect the cable quality from different perspectives, avoiding subjective biases as much as possible. This ensures that the selected indicators meet both theoretical and practical requirements, making them applicable in the field of quality evaluation. Here, we take low-voltage power cables as an example. The $20^\circ\mathrm{C}$ conductor DC resistance, minimum insulation thickness, average insulation thickness, and minimum sheath thickness are selected as performance indicators. The size of the conductor DC resistance directly affects the cable's current-carrying capacity. The thickness and uniformity of the insulation and sheath impact the product's lifespan and can reflect the level of production processes and equipment control to some extent.
	
	Based on the low-voltage power cable inspection report data from ECP, the nonconformity rates caused by the four indicators mentioned above are calculated. A judgment matrix is constructed, and through the Analytic Hierarchy Process, comprehensive evaluations are conducted to determine the weight values corresponding to the four indicators. The obtained weights are shown in Table \ref{weight}.

	\begin{table}[h]
		\newcommand{\tabincell}[2]{\begin{tabular}{@{}#1@{}}#2\end{tabular}}
		\renewcommand\tabcolsep{5.0pt}
		\renewcommand{\arraystretch}{1.2}
		\caption{The weight value of the performance index calculated by the AHP}
		\centering
		\resizebox{\columnwidth}{!}{
			\begin{tabular}{c|c|c}
				\hline
				\multirow{1}{*}{\textbf{Serial Number}}&\multirow{1}{*}{\textbf{Name of Indicator}}&\multirow{1}{*}{\textbf{Weight $W_i$}}\\
				\hline
				\multirow{1}{*}{1}&conductor DC resistance&0.321\\
				\hline
				\multirow{1}{*}{2}&minimum insulation thickness&0.214\\
				\hline
				\multirow{1}{*}{3}&average insulation thickness&0.222\\
				\hline
				\multirow{1}{*}{4}&minimum sheath thickness&0.243\\
				\hline
			\end{tabular}
		}
		\label{weight}
	\end{table}

	\begin{table*}[ht]
		\newcommand{\tabincell}[2]{\begin{tabular}{@{}#1@{}}#2\end{tabular}}
		\renewcommand\tabcolsep{5.0pt}
		\renewcommand{\arraystretch}{1.2}
		\caption{Guidelines for Classification and Grading of Power Cable Quality}
		\centering
		\resizebox{\textwidth}{!}{
			\begin{tabular}{c|ccccc}
				\hline
				\multicolumn{2}{c|}{\multirow{2}{*}{Items}} & \multicolumn{1}{c|}{Excellent} & \multicolumn{1}{c|}{Good} & \multicolumn{1}{c|}{Qualified} & \multicolumn{1}{c}{Basic Qualified}\\
				\multicolumn{2}{c|}{} & \multicolumn{1}{c|}{($G_i=a$)} & \multicolumn{1}{c|}{($G_i=b$)} & \multicolumn{1}{c|}{($G_i=c$)} & \multicolumn{1}{c}{($G_i=d$)}\\
				\hline
				\multicolumn{1}{c|}{\multirow{3}{*}{DC resistance of the conductor}} & \multicolumn{1}{c|}{Copper core} & \multicolumn{1}{c|}{$d \leq -5$} & \multicolumn{1}{c|}{$-5 < \delta \leq -3$} & \multicolumn{1}{c|}{$-3 < \delta \leq -1$} & \multicolumn{1}{c}{\multirow{3}{*}{$-1 < \delta \leq 0$}}\\
				\cline{2-5}
				& \multicolumn{1}{c|}{Aluminum core} & \multicolumn{1}{c|}{\multirow{2}{*}{$\delta \leq -7$}} & \multicolumn{1}{c|}{\multirow{2}{*}{$-7 < \delta \leq -4$}} & \multicolumn{1}{c|}{\multirow{2}{*}{$-4 < \delta \leq -1$}} & \multicolumn{1}{c}{}\\
				\cline{2-2}
				& \multicolumn{1}{c|}{Aluminum alloy core} & \multicolumn{1}{c|}{} & \multicolumn{1}{c|}{} & \multicolumn{1}{c|}{} & \multicolumn{1}{c}{}\\
				\hline
				\multicolumn{2}{c|}{\multirow{1}{*}{average insulation thickness}} & \multicolumn{1}{c|}{$\delta \geq 25$} & \multicolumn{1}{c|}{$15 \leq \delta < 25$} & \multicolumn{1}{c|}{$5 \leq \delta < 15$} & \multicolumn{1}{c}{$0 \leq \delta < 5$}\\
				\hline
				\multicolumn{2}{c|}{\multirow{1}{*}{minimum insulation thickness}} & \multicolumn{1}{c|}{$\delta \geq 25$} & \multicolumn{1}{c|}{$15 \leq \delta < 25$} & \multicolumn{1}{c|}{$5 \leq \delta < 15$} & \multicolumn{1}{c}{$0 \leq \delta < 5$}\\
				\hline
				\multicolumn{2}{c|}{\multirow{1}{*}{minimum thickness of the sheath}} & \multicolumn{1}{c|}{$\delta \geq 50$} & \multicolumn{1}{c|}{$30 \leq \delta < 50$} & \multicolumn{1}{c|}{$15 \leq \delta < 30$} & \multicolumn{1}{c}{$0 \leq \delta < 15$}\\
				\hline
		\end{tabular}}
		\label{Guidelines}
	\end{table*}
	
	\begin{table*}[t]
		\newcommand{\tabincell}[2]{\begin{tabular}{@{}#1@{}}#2\end{tabular}}
		\renewcommand{\arraystretch}{2.0}
		\renewcommand\tabcolsep{5.0pt}
		\caption{Example of Weighted Quality Score for Power Cables}
		\centering
		\resizebox{\textwidth}{!}{
			\begin{tabular}{c|ccccccccccccc}
				\hline
				\multicolumn{1}{c|}{\multirow{3}{*}{\textbf{Serial Number}}} & \multicolumn{3}{c|}{\textbf{DC resistance of the conductor}} & \multicolumn{3}{c|}{\textbf{minimum insulation thickness}} & \multicolumn{3}{c|}{\textbf{average insulation thickness}} & \multicolumn{3}{c|}{\textbf{minimum thickness of the sheath}} & \multicolumn{1}{c}{\multirow{3}{*}{\textbf{Quality Score $M$}}}\\
				&  \multicolumn{3}{c|}{($W_1=0.321$)} & \multicolumn{3}{c|}{($W_2=0.214$)} & \multicolumn{3}{c|}{($W_3=0.222$)} & \multicolumn{3}{c|}{($W_4=0.243$)} &\\
				\cline{2-13}
				& \multicolumn{1}{c|}{Deviation $\delta$} & \multicolumn{1}{c|}{Level} & \multicolumn{1}{c|}{Score $G_1$}
				& \multicolumn{1}{c|}{Deviation $\delta$} & \multicolumn{1}{c|}{Level} & \multicolumn{1}{c|}{Score $G_2$}
				& \multicolumn{1}{c|}{Deviation $\delta$} & \multicolumn{1}{c|}{Level} & \multicolumn{1}{c|}{Score $G_3$}
				& \multicolumn{1}{c|}{Deviation $\delta$} & \multicolumn{1}{c|}{Level} & \multicolumn{1}{c|}{Score $G_4$} &\\
				\cline{1-14}
				\multicolumn{1}{c|}{1}
				& \multicolumn{1}{c|}{-2.4} & \multicolumn{1}{c|}{Qualified} & \multicolumn{1}{c|}{75}
				& \multicolumn{1}{c|}{9.8} & \multicolumn{1}{c|}{Qualified} & \multicolumn{1}{c|}{75}
				& \multicolumn{1}{c|}{6.9} & \multicolumn{1}{c|}{Qualified} & \multicolumn{1}{c|}{75}
				& \multicolumn{1}{c|}{38.3} & \multicolumn{1}{c|}{Qualified} & \multicolumn{1}{c|}{85}
				&\multicolumn{1}{c}{77.43}\\
				\hline
				\multicolumn{1}{c|}{2}
				& \multicolumn{1}{c|}{-1.0} & \multicolumn{1}{c|}{Qualified} & \multicolumn{1}{c|}{75}
				& \multicolumn{1}{c|}{12.3} & \multicolumn{1}{c|}{Qualified} & \multicolumn{1}{c|}{75}
				& \multicolumn{1}{c|}{4.2} & \multicolumn{1}{c|}{Basic Qualified} & \multicolumn{1}{c|}{60}
				& \multicolumn{1}{c|}{44.4} & \multicolumn{1}{c|}{Good} & \multicolumn{1}{c|}{85}
				&\multicolumn{1}{c}{74.1}\\
				\hline
				\multicolumn{1}{c|}{3}
				& \multicolumn{1}{c|}{-20.9} & \multicolumn{1}{c|}{Excellent} & \multicolumn{1}{c|}{100}
				& \multicolumn{1}{c|}{36.9} & \multicolumn{1}{c|}{Excellent} & \multicolumn{1}{c|}{100}
				& \multicolumn{1}{c|}{33.7} & \multicolumn{1}{c|}{Excellent} & \multicolumn{1}{c|}{100}
				& \multicolumn{1}{c|}{11.5} & \multicolumn{1}{c|}{Basic Qualified} & \multicolumn{1}{c|}{60}
				&\multicolumn{1}{c}{90.28}\\
				\hline
				\multicolumn{1}{c|}{4}
				& \multicolumn{1}{c|}{-1.2} & \multicolumn{1}{c|}{Qualified} & \multicolumn{1}{c|}{75}
				& \multicolumn{1}{c|}{37.5} & \multicolumn{1}{c|}{Excellent} & \multicolumn{1}{c|}{100}
				& \multicolumn{1}{c|}{33.3} & \multicolumn{1}{c|}{Excellent} & \multicolumn{1}{c|}{100}
				& \multicolumn{1}{c|}{61.1} & \multicolumn{1}{c|}{Excellent} & \multicolumn{1}{c|}{100}
				&\multicolumn{1}{c}{91.975}\\
				\hline
				\multicolumn{1}{c|}{5}
				& \multicolumn{1}{c|}{1.5} & \multicolumn{1}{c|}{Qualified} & \multicolumn{1}{c|}{75}
				& \multicolumn{1}{c|}{27.3} & \multicolumn{1}{c|}{Excellent} & \multicolumn{1}{c|}{100}
				& \multicolumn{1}{c|}{30.6} & \multicolumn{1}{c|}{Excellent} & \multicolumn{1}{c|}{100}
				& \multicolumn{1}{c|}{19.0} & \multicolumn{1}{c|}{Qualified} & \multicolumn{1}{c|}{75}
				&\multicolumn{1}{c}{85.9}\\
				\hline
				\multicolumn{1}{c|}{6}
				& \multicolumn{1}{c|}{-1.6} & \multicolumn{1}{c|}{Qualified} & \multicolumn{1}{c|}{75}
				& \multicolumn{1}{c|}{107.9} & \multicolumn{1}{c|}{Excellent} & \multicolumn{1}{c|}{100}
				& \multicolumn{1}{c|}{107.1} & \multicolumn{1}{c|}{Excellent} & \multicolumn{1}{c|}{100}
				& \multicolumn{1}{c|}{46.9} & \multicolumn{1}{c|}{Good} & \multicolumn{1}{c|}{85}
				&\multicolumn{1}{c}{88.33}\\
				\hline
				\multicolumn{1}{c|}{7}
				& \multicolumn{1}{c|}{-3.5} & \multicolumn{1}{c|}{Good} & \multicolumn{1}{c|}{85}
				& \multicolumn{1}{c|}{25.0} & \multicolumn{1}{c|}{Excellent} & \multicolumn{1}{c|}{100}
				& \multicolumn{1}{c|}{75} & \multicolumn{1}{c|}{Qualified} & \multicolumn{1}{c|}{75}
				& \multicolumn{1}{c|}{60} & \multicolumn{1}{c|}{Basic Qualified} & \multicolumn{1}{c|}{60}
				&\multicolumn{1}{c}{79.915}\\
				\hline
				\multicolumn{1}{c|}{8}
				& \multicolumn{1}{c|}{-10.2} & \multicolumn{1}{c|}{Excellent} & \multicolumn{1}{c|}{100}
				& \multicolumn{1}{c|}{63.5} & \multicolumn{1}{c|}{Excellent} & \multicolumn{1}{c|}{100}
				& \multicolumn{1}{c|}{50.0} & \multicolumn{1}{c|}{Excellent} & \multicolumn{1}{c|}{100}
				& \multicolumn{1}{c|}{16.7} & \multicolumn{1}{c|}{Qualified} & \multicolumn{1}{c|}{75}
				&\multicolumn{1}{c}{93.925}\\
				\hline
				\multicolumn{1}{c|}{9}
				& \multicolumn{1}{c|}{-8.2} & \multicolumn{1}{c|}{Excellent} & \multicolumn{1}{c|}{100}
				& \multicolumn{1}{c|}{26.7} & \multicolumn{1}{c|}{Excellent} & \multicolumn{1}{c|}{100}
				& \multicolumn{1}{c|}{20.0} & \multicolumn{1}{c|}{Good} & \multicolumn{1}{c|}{85}
				& \multicolumn{1}{c|}{9.9} & \multicolumn{1}{c|}{Basic Qualified} & \multicolumn{1}{c|}{60}
				&\multicolumn{1}{c}{86.95}\\
				\hline
				\multicolumn{1}{c|}{10}
				& \multicolumn{1}{c|}{-3.6} & \multicolumn{1}{c|}{Good} & \multicolumn{1}{c|}{85}
				& \multicolumn{1}{c|}{37.0} & \multicolumn{1}{c|}{Excellent} & \multicolumn{1}{c|}{100}
				& \multicolumn{1}{c|}{27.8} & \multicolumn{1}{c|}{Excellent} & \multicolumn{1}{c|}{100}
				& \multicolumn{1}{c|}{29.4} & \multicolumn{1}{c|}{Qualified} & \multicolumn{1}{c|}{75}
				&\multicolumn{1}{c}{89.11}\\
				\hline
		\end{tabular}}
		\label{score}
	\end{table*}
	
	According to Table \ref{weight}, the weight distribution of performance indicators reveals that the insulation structural performance accounts for 43.6\%, the electrical performance accounts for 32.1\%, and the sheath structural performance accounts for 24.3\%. This effectively demonstrates the characteristics of low-voltage power cables. Among these, the insulation structure is a critical component of the cable, and its thickness is positively correlated with the electrical strength. If the thickness is too thin, it can lead to cable breakdown and short circuits, making the insulation structure highly significant for cable quality. Additionally, with the rise in the price of raw materials such as copper and aluminum, some manufacturers have resorted to reducing materials in cable conductors. Consequently, the failure rate of the DC resistance test conducted at $20^\circ\mathrm{C}$ has been increasing in recent years. This test has garnered attention and has become a crucial indicator for assessing cable quality. The sheath, serving as the cable's outer layer, plays a crucial role in its protection. The thickness of the sheath is directly related to its protective function. In conclusion, the weighted results not only align with expert expectations but also consider historical testing results. Higher weights are allocated to items with lower qualification rates. Therefore, the weight allocation results are reasonable and accurately reflect the quality characteristics of cable products.
	
	\subsection{Classification of test result levels}

	Since the evaluation standard values of cable products of different specifications are inconsistent, the deviation d between the measured values and the standard values is calculated as the basis for the division when the guidelines for the division of detection grades are formulated to take into account the difference in standard values between products of various specifications. The calculation formula for the deviation $\delta$ is as follows
	
	\begin{footnotesize}
		\begin{equation}
			\begin{aligned}
				\delta = \frac{X - Y}{Y},
			\end{aligned}
		\end{equation}
	\end{footnotesize}
	
	\noindent where $X$ denotes the measured value and $Y$ denotes the standard value. The qualified products in this paper are further divided into four levels: Excellent, Good, Qualified, and Basic qualified. Each level is assigned a corresponding score: $a=100$, $b=85$, $c=75$, $d=60$. The classification guidelines are presented in Table \ref{Guidelines}.

	\subsection{Calculating the quality-weighted scores and sampling}
	
	According to the grading guidelines of power cables, the performance data of the inspected products are individually scored. By combining the indicator weight values and using the quality weighting formula as Eq. \ref{qualityWeight}, the quality weighted scores ($M$) for the sampled products are calculated. Taking 10 samples of power cables for inspection as an example, the results of the grading and the calculation of the quality weighted scores ($M$) for each indicator are shown in Table \ref{score}. Sort the items in ascending order based on their comprehensive performance values. Assuming a sampling quantity of 2, select material/equipment numbers 2 and 1 for sampling and testing.
	
    \subsection{Model Comparison}
To evaluate the effectiveness of our proposed method, we compare it with the Artificial Neural Network(ANN)\cite{LI2020108912} and Random Forest (RF)\cite{CHUN2020119238} on the level prediction of power cable quality. The ANN model is a type of multilayer feedforward neural network which can efficiently and accurately obtain the mathematical relationship between input and output variables. RF is an ensemble model that combines multiple decision trees to improve the stability and accuracy of prediction. In addition, RF can estimate the importance of each variable in the prediction and management of learning situations
with small sample sizes without overfitting problems. Therefore, we choose to compare with these two methods. From the results in Table \ref{comparison}, we can see that our method has a significant performance advantage over the compared approaches. Of all the products, products 5 and 10 are close to the nonconforming level and should be selected for testing. Compared with the compared methods, only our method can accurately predict the quality levels of 5 and 10, which indicates that our sampling method can accurately sample representative products that reflect the overall quality.

\begin{table}[h]
		\newcommand{\tabincell}[2]{\begin{tabular}{@{}#1@{}}#2\end{tabular}}
		\renewcommand\tabcolsep{5.0pt}
		\renewcommand{\arraystretch}{1.2}
        \caption{Results on the level prediction of power cable quality}
        \centering
        \resizebox{\columnwidth}{!}
        {

            \begin{threeparttable}
            \resizebox{\linewidth}{!}{
                \begin{tabular}{c|c|c|c|c}
                  \hline
                  \textbf{Serial Number} & \textbf{Level} & \textbf{ANN} & \textbf{Random Forest}& \textbf{Ours}\\
                  \hline
                  1 & Excellent& Qualified  &Excellent& Qualified\\
                  2 & Qualified& Qualified  &Basic Qualified& Qualified\\
                  3 & Qualified& Excellent & Excellent& Qualified\\
                  4 & Excellent& Qualified &Excellent & Excellent\\
                  5 & Basic Qualified& Qualified &Good&   Basic Qualified\\
                  6 & Excellent& Qualified &Excellent & Excellent\\
                  7 & Excellent& Good  & Excellent& Excellent\\
                  8 & Qualified& Excellent & Excellent& Good\\
                  9 & Excellent& Excellent & Basic Qualified &Qualified\\
                  10 & Basic Qualified&Good & Qualified&Basic Qualified\\
                  \hline
                \end{tabular}
                }

          \end{threeparttable}
          }
        \label{comparison}
        \end{table}

	\section{Conclusions\label{conclusion}}

This paper addresses the practical requirements for material procurement sampling in distribution network systems and presents an enhanced sampling model based on the AHP. The approach leverages a substantial amount of test and inspection data from the smart supply chain system of the State Grid Corporation to assign quality scores to sampled material equipment. By prioritizing the quality scores, the model strategically selects sampled material equipment for testing according to the specified sampling quantity, thereby enhancing the accuracy of inspections compared to random sampling methods. The proposed material testing sampling model in this paper considers scientific rigor, feasibility, and the cost of sampling work, with the aim of improving both quality and efficiency. It can be directly applied to the ongoing critical material procurement sampling efforts in distribution networks, providing assurance for quality control before equipment installation.
	
	\section{Data Availability}
	The data used to support the study are included in the paper.
	
	\section{Conflicts of Interest}
	The authors declare that there is no conflict of interest regarding the publication of this paper.

	\bibliographystyle{unsrt}
	\bibliography{Paper}

\begin{thebibliography}{10}

\bibitem{li2017research}
Na~Li, Zhenhua Zhu, Ming Li, Ying Lin, Xiaoliang Wang, and Qinglong Liu.
\newblock Research on reliability evaluation of power system including improved
  monte carlo and parallel calculation.
\newblock In {\em 2017 2nd International Conference on Power and Renewable
  Energy (ICPRE)}, pages 534--538. IEEE, 2017.

\bibitem{weaver2016cyber}
Gabriel~A Weaver, Kate Davis, Charles~M Davis, Edmond~J Rogers, Rakesh~B Bobba,
  Saman Zonouz, Robin Berthier, Peter~W Sauer, and David~M Nicol.
\newblock Cyber-physical models for power grid security analysis: 8-substation
  case.
\newblock In {\em 2016 IEEE International Conference on Smart Grid
  Communications (SmartGridComm)}, pages 140--146. IEEE, 2016.

\bibitem{saaty1988analytic}
Thomas~L Saaty.
\newblock What is the analytic hierarchy process?
\newblock In {\em Mathematical Models for Decision Support}, volume~48, pages
  109--121, Berlin, Heidelberg, 1988. Springer Berlin Heidelberg.

\bibitem{darko2019review}
Amos Darko, Albert Ping~Chuen Chan, Ernest~Effah Ameyaw, Emmanuel~Kingsford
  Owusu, Erika Pärn, and David~John Edwards.
\newblock Review of application of analytic hierarchy process (\uppercase{AHP})
  in construction.
\newblock {\em International Journal of Construction Management},
  19(5):436--452, 2019.

\bibitem{saaty1986absolute}
Thomas~L Saaty.
\newblock Absolute and relative measurement with the ahp. the most livable
  cities in the united states.
\newblock {\em Socio-Economic Planning Sciences}, 20(6):327--331, 1986.

\bibitem{de2013absolute}
Fabio~De Felice and Antonella Petrillo.
\newblock Absolute measurement with analytic hierarchy process: A case study
  for italian racecourse.
\newblock {\em International Journal of Applied Decision Sciences},
  6(3):209--227, 2013.

\bibitem{salomon2016absolute}
Valerio A.~P. Salomon.
\newblock Absolute measurement and ideal synthesis on \uppercase{AHP}.
\newblock {\em International Journal of the Analytic Hierarchy Process},
  8(3):538--545, 2016.

\bibitem{WU2014239}
Haizhen Wu and Kondaswamy Govindaraju.
\newblock Computer-aided variables sampling inspection plans for compositional
  proportions and measurement error adjustment.
\newblock {\em Computers and Industrial Engineering}, 72:239--246, 2014.

\bibitem{SOMMER201989}
Andreas Sommer and Ansgar Steland.
\newblock Multistage acceptance sampling under nonparametric dependent sampling
  designs.
\newblock {\em Journal of Statistical Planning and Inference}, 199:89--113,
  2019.

\bibitem{BANIHASHEMI2021107155}
Atefe Banihashemi, Mohammad Saber~Fallah Nezhad, and Amirhossein Amiri.
\newblock A new approach in the economic design of acceptance sampling plans
  based on process yield index and taguchi loss function.
\newblock {\em Computers and Industrial Engineering}, 159:107155, 2021.

\bibitem{thomas2023design}
Julia~Thampy Thomas and Mahesh Kumar.
\newblock Design of generalized fuzzy multiple deferred state
  (\uppercase{GFMDS}) sampling plan for attributes.
\newblock {\em arXiv preprint arXiv:2302.01606}, 2023.

\bibitem{do2013calculating}
Huu~Tuan Do, Shang-Lien Lo, and Lan~Anh Phan~Thi.
\newblock Calculating of river water quality sampling frequency by the analytic
  hierarchy process (\uppercase{AHP}).
\newblock {\em Environmental monitoring and assessment}, 185:909--916, 2013.

\bibitem{apostolopoulos2016regional}
Nikolaos Apostolopoulos and Panagiotis Liargovas.
\newblock Regional parameters and solar energy enterprises: Purposive sampling
  and group ate \uppercase{AHP} approach.
\newblock {\em International Journal of Energy Sector Management},
  10(1):19--37, 2016.

\bibitem{kumar2017soil}
Nirmal Kumar, SK~Singh, VN~Mishra, GP~Reddy, and RK~Bajpai.
\newblock Soil quality ranking of a small sample size using \uppercase{AHP}.
\newblock {\em Journal of Soil and Water Conservation}, 16(4):339--346, 2017.

\bibitem{singh2023assessing}
Rupanjali Singh, CB~Majumder, and Ajit~Kumar Vidyarthi.
\newblock Assessing the impacts of industrial wastewater on the inland surface
  water quality: an application of analytic hierarchy process (\uppercase{AHP})
  model-based water quality index and \uppercase{GIS} techniques.
\newblock {\em Physics and Chemistry of the Earth, Parts A/B/C},
  129:103314(1)--103314(14), 2023.

\bibitem{VAIDYA20061}
Omkarprasad~S. Vaidya and Sushil Kumar.
\newblock Analytic hierarchy process: \uppercase{A}n overview of applications.
\newblock {\em European Journal of Operational Research}, 169(1):1--29, 2006.

\bibitem{Alireza2009}
Alireza Davoodi.
\newblock On inconsistency of a pairwise comparison matrix.
\newblock {\em International Journal of Industrial Mathematics}, 1:343--350,
  2009.

\bibitem{AGUARON2003137}
Juan Aguarón and José~Marı́a Moreno-Jiménez.
\newblock The geometric consistency index: \uppercase{A}pproximated thresholds.
\newblock {\em European Journal of Operational Research}, 147(1):137--145,
  2003.

\bibitem{SHRESTHA2004185}
Ram~K Shrestha, Janaki~R.R Alavalapati, and Robert~S Kalmbacher.
\newblock Exploring the potential for silvopasture adoption in south-central
  \uppercase{F}lorida: an application of \uppercase{SWOT}–\uppercase{AHP}
  method.
\newblock {\em Agricultural Systems}, 81(3):185--199, 2004.

\bibitem{FRANEK2014164}
Jiří Franek and Aleš Kresta.
\newblock Judgment scales and consistency measure in \uppercase{AHP}.
\newblock {\em Procedia Economics and Finance}, 12:164--173, 2014.

\bibitem{5579387}
Bai Hanbin and Wang Nuanchen.
\newblock Research on the selection of scale in \uppercase{AHP}.
\newblock In {\em 2010 3rd International Conference on Advanced Computer Theory
  and Engineering(ICACTE)}, volume~6, pages V6--108--V6--111, Chengdu, China,
  2010.

\bibitem{LI2020108912}
Yang Li, Bo~Yu, Baicun Wang, Tae~Hwa Lee, and Mihaela Banu.
\newblock Online quality inspection of ultrasonic composite welding by
  combining artificial intelligence technologies with welding process
  signatures.
\newblock {\em Materials $\&$ Design}, 194:108912(1)--108912(10), 2020.

\bibitem{CHUN2020119238}
Pang jo~Chun, Isao Ujike, Kohei Mishima, Masahiro Kusumoto, and Shinichiro
  Okazaki.
\newblock Random forest-based evaluation technique for internal damage in
  reinforced concrete featuring multiple nondestructive testing results.
\newblock {\em Construction and Building Materials}, 253:119238(1)--119238(11),
  2020.

\end{thebibliography}

\end{document}